\documentclass[conference, letterpaper]{IEEEtran}

\usepackage{balance}
\usepackage{fancyhdr}
\usepackage{cite}
\usepackage{amsmath,color}
\usepackage{amssymb}
\usepackage{latexsym}
\usepackage{longtable}
\usepackage{mathrsfs}
\usepackage{color}
\usepackage{multicol}
\usepackage{amsfonts}
\usepackage{array}
\usepackage[nomain,acronym,automake,nopostdot]{glossaries}
\definecolor{sred}{RGB}{200,21,0}
\definecolor{sblue}{RGB}{0,51,160}
\usepackage[a4 paper, top=0.75in, bottom=1in, left=0.8in, right=0.8in]{geometry}

\ifCLASSINFOpdf
   \usepackage[pdftex]{graphicx}
   \graphicspath{{../pdf/}{../jpeg/}}
   \DeclareGraphicsExtensions{.pdf,.jpeg,.png}
\else
   \usepackage[dvips]{graphicx}
   \graphicspath{{../eps/}}
   \DeclareGraphicsExtensions{.eps}
\fi

\pdfoutput=1

\newtheorem{prop}{Proposition}
\newtheorem{defn}{Definition}

\newcommand{\matc}[1]{\mbox{\boldmath $\mathcal{#1}$}}

\newcommand{\figref}[1]{Fig. \ref{#1}}

\usepackage{bbm}
\usepackage{setspace}
\usepackage[ruled]{algorithm2e}
\usepackage{amsmath}
\makeglossaries
\newacronym{ils}{ILS}{integer least-squares}
\newacronym{snr}{SNR}{signal-to-noise ratios}
\newacronym{wl}{WL}{widely linear}
\newacronym{mlsd}{MLSD}{maximum-likelihood sequence detection}
\newacronym{map}{MAP}{maximum a posteriori}
\newacronym{lmmse}{LMMSE}{linear minimum mean-square error}
\newacronym{ut}{UT}{user terminal}
\newacronym{bs}{BS}{base station}
\newacronym{ann}{ANN}{artificial neural network}
\newacronym{dnn}{DNN}{deep neural network}
\newacronym{snn}{SNN}{super neural networks}
\newacronym{gmnn}{G-MNN}{giant modular neural network}
\newacronym{mnn}{MNN}{modular neural network}
\newacronym{pic}{PIC}{parallel interference cancellation}
\newacronym{siso}{SISO}{single-input single-output}
\newacronym{mimo}{MIMO}{multiple-input multiple-output}
\newacronym{mud}{MUD}{multiuser detection}
\newacronym{amp}{AMP}{approximate message passing}
\newacronym{zf}{ZF}{zero forcing}
\newacronym{mf}{MF}{matched filter}
\newacronym{cdma}{CDMA}{code division multiple access}
\newacronym{noma}{NOMA}{non-orthogonal multiple access}
\newacronym{csir}{CSIR}{channel state information at the receiver}
\newacronym{sic}{SIC}{successive interference cancellation}
\newacronym{cdf}{CDF}{cumulative distribution function}
\newacronym{ncl}{NC-Learning}{non-coherent learning}
\newacronym{cel}{CE-Learning}{channel equalized learning}
\newacronym{dcl}{DC-Learning}{direct coherent learning}
\newacronym{bp}{BP}{back propagation}
\newacronym{vq}{VQ}{vector quantization}
\newacronym{awgn}{AWGN}{additive white Gaussian noise}
\newacronym{ai}{AI}{artificial intelligence}
\newacronym{dsp}{DSP}{digital signal processing}
\newacronym{csi}{CSI}{channel state information}
\newacronym{nn}{NN}{nearest-neighbor}
\newacronym{clnn}{CL-NN}{cluster-level nearest neighbor}
\newacronym{clknn}{CL-kNN}{cluster-level kNN}
\newacronym{ber}{BER}{bit error rate}
\newacronym{sgd}{SGD}{stochastic gradient descent}
\newacronym{osgd}{O-SGD}{orthogonal stochastic gradient descent}
\newacronym{ml}{ML}{machine learning}
\newacronym{rls}{RLS}{recursive least-square}
\begin{document}

\title{\huge An Orthogonal-SGD based Learning Approach for MIMO Detection under Multiple Channel Models}

\author{\IEEEauthorblockN{Songyan Xue, Yi Ma, and Rahim Tafazolli}
\IEEEauthorblockA{Institute for Communication Systems (ICS),
University of Surrey, Guildford, England, GU2 7XH\\
E-mail: (songyan.xue, y.ma, r.tafazolli)@surrey.ac.uk}}
\markboth{}%
{}
\maketitle

\begin{abstract}
In this paper,  an orthogonal stochastic gradient descent (O-SGD) based learning approach is proposed to tackle the wireless channel over-training problem inherent in artificial neural network (ANN)-assisted MIMO signal detection. Our basic idea lies in the discovery and exploitation of the training-sample orthogonality between the current training epoch and past training epochs. Unlike the conventional SGD that updates the neural network simply based upon current training samples, O-SGD discovers the correlation between current training samples and historical training data, and then updates the neural network with those uncorrelated components. The network updating occurs only in those identified null subspaces. By such means, the neural network can understand and memorize uncorrelated components between different wireless channels, and thus is more robust to wireless channel variations. This hypothesis is confirmed through our extensive computer simulations as well as performance comparison with the conventional SGD approach.
\end{abstract}

\IEEEpeerreviewmaketitle

\section{Introduction}\label{sec1}
Detection of \gls{mimo} signals through \gls{ml} has demonstrated remarkable advantages in terms of their strong parallel-processing ability, good performance-complexity tradeoff, as well as self-optimization with respect to the dynamics of wireless channels \cite{xue2019learn}. More remarkably, data-driven \gls{ml} approaches are model independent, i.e., they learn to detect signals without the need of an explicit model of the signal propagation (e.g. \cite{8437142,DBLP:journals/corr/OSheaEC17,8761999}). This is particularly useful for receivers to reconstruct signals from random nonlinear distortions, which are often very hard to handle with hand-engineered approaches. Meanwhile, \gls{ml}-assisted wireless receivers can also be model-driven (e.g. \cite{8227772,8646357,chen2019efficient,mohammad2019complexityscalable}), which can take advantage of the model knowledge to mitigate the {\em curse of dimensionality} problem inherent in the deep learning procedure. Moreover, \gls{ml} and hand-engineered approaches can work together to form a synergy when conducting the signal detection \cite{khani2019adaptive,8589084,8755566, 2018arXiv180401002T,8761049,8613336,nguyen2019deep}. 

Despite already numerous contributions in this domain, there are very few results that have been reported so far, concerning the wireless channel over-training problem. More specifically, current \gls{ml}-assisted receivers are trained mainly for a specific channel model; such as the \gls{mimo} Rayleigh-fading channel. However, a receiver that is well trained for one channel model is often too sub-optimum or even unsuitable for other channel models. This is also known as the training set over-fitting problem in the general artificial intelligence domain. In the literature, there are a couple of ways to handle the over-training problem. One approach is called continual learning, which aims to inject new knowledge without forgetting previously learned knowledge.  As a consequence, machines will always adapt themselves to be better optimized for latest training samples (i.e. new channel models in telecommunications) \cite{toneva2018an}. The other approach is called multi-task learning \cite{DBLP:journals/corr/ZhangY17aa} which aims to improve all training tasks simultaneously by combining their common features. These approaches have already achieved promising results in traditional \gls{ml} applications, such as natural language processing or image/video recognition; however, it is still not clear whether these approaches can be cost-effective to handle wireless channels that are random, continuous, and infinite in their states. 

In this paper, we introduce an \gls{osgd} algorithm to tackle the wireless channel over-training problem when machines learn to detect communication signals in \gls{mimo} fading channels \footnote{\gls{osgd} is suitable for detecting communication signals in general cases. This paper focuses on the \gls{mimo} signal model for the sake of concise presentation and clear concept delivery.}. The basic idea lies in the discovery and exploitation of the orthogonality of training samples between the current training epoch and past training epochs. More specifically, the \gls{osgd} algorithm does not update the neural network simply based upon training samples of the current epoch. Instead, it first discovers the correlation between current training samples and historical training data, and then update the neural network with those uncorrelated components. The network updating occurs only in those identified null subspaces. By such means, the neural network can understand and memorize uncorrelated components between different training tasks (e.g. channel models). This idea is evaluated for \gls{ann}-assisted \gls{mimo} detection with various channel models. It is shown, through computer simulations, that \gls{osgd} is very robust to channel model variations as well as SNR variations.

\section{System model and Preliminaries}\label{sec2}
\subsection{MIMO System Model and Optimum Detection}\label{sec2a}
Consider \gls{mimo} uplink communications, where $M$ transmit antennas simultaneously talking with $N$ receive antennas through the wireless channel ($N\geq M$). It is also assumed that the receive antennas can fully cooperate and share their received waveform for joint signal processing. The discrete-time equivalent baseband signal model can be described as the following matrix form
\begin{equation}\label{eq01}
\mathbf{y}=\mathbf{H}\mathbf{x}+\mathbf{v}
\end{equation}
where $\mathbf{y}=[y_0,\dots,y_{N-1}]^T$ stands for the spatial-domain received signal block, $\mathbf{x}=[x_0,\dots,x_{M-1}]^T$ for the transmitted signal block with zero mean and identical covariance $\sigma_x^2$. Each symbol of $\mathbf{x}$ is independently drawn from a finite-alphabet set $\matc{A}$ with $\left | \matc{A} \right |=L$, $\mathbf{H}\in \mathbb{C}^{N\times M}$ for the \gls{mimo} channel matrix, and $\mathbf{v}$ for the additive white Gaussian noise (AWGN) with $\mathbf{v}\sim CN(0,\sigma _v^2\mathbf{I})$. Moreover, the superscript $[\cdot]^T$ stands for the vector/matrix transpose, and $\mathbf{I}$ for the identity matrix.

The general problem of \gls{mimo} signal detection is to form the decision $\hat{\mathbf{x}}$ based upon the received signal $\mathbf{y}$ and channel matrix $\mathbf{H}$. The maximum-likelihood solution (or equivalently the sphere decoding) is \gls{ils} optimum by achieving the following objective function
\begin{equation}\label{eq02}
\hat{\mathbf{x}}=\underset{\mathbf{x}\in \matc{A}^M}{\arg \min} \left \|\mathbf{y}-\mathbf{Hx}  \right \|^2
\end{equation}
where $\left \| \cdot \right \|$ stands for the Euclidean norm.

\subsection{On Scalability of \gls{ann}-assisted MIMO Detection}\label{sec2b}
In the \gls{ml} theory, the \gls{ils} problem in \eqref{eq02} is actually a Bayesian optimization problem. The basic principle can be described as:

\begin{prop}[See \cite{8761999}]\label{prop1}
Given the \gls{mimo} channel matrix $\mathbf{H}$ and the finite-alphabet set $\mathbf{x}\in\matc{A}^M$, \gls{ml} is able to establish the link between $\mathbf{y}$ and $\mathbf{H}\mathbf{x}$ according to the maximum {\em a} posteriori probability $p(\mathbf{H}\mathbf{x}|\mathbf{y})$.
\end{prop}

If the \gls{mimo} channel $\mathbf{H}$ is fixed, we have the maximum {\em a} posteriori probability $p(\mathbf{H}\mathbf{x}|\mathbf{y}) = p(\mathbf{x}|\mathbf{y})$. The size of the finite-alphabet set grows exponentially with the number of transmit antennas and polynomially with the modulation order. Despite very high complexities for a large-MIMO, the \gls{ml} procedure can be managed by employing the high-performance parallel computing technology.

\begin{figure}[t]
\begin{center}
\includegraphics[width=0.9\columnwidth]{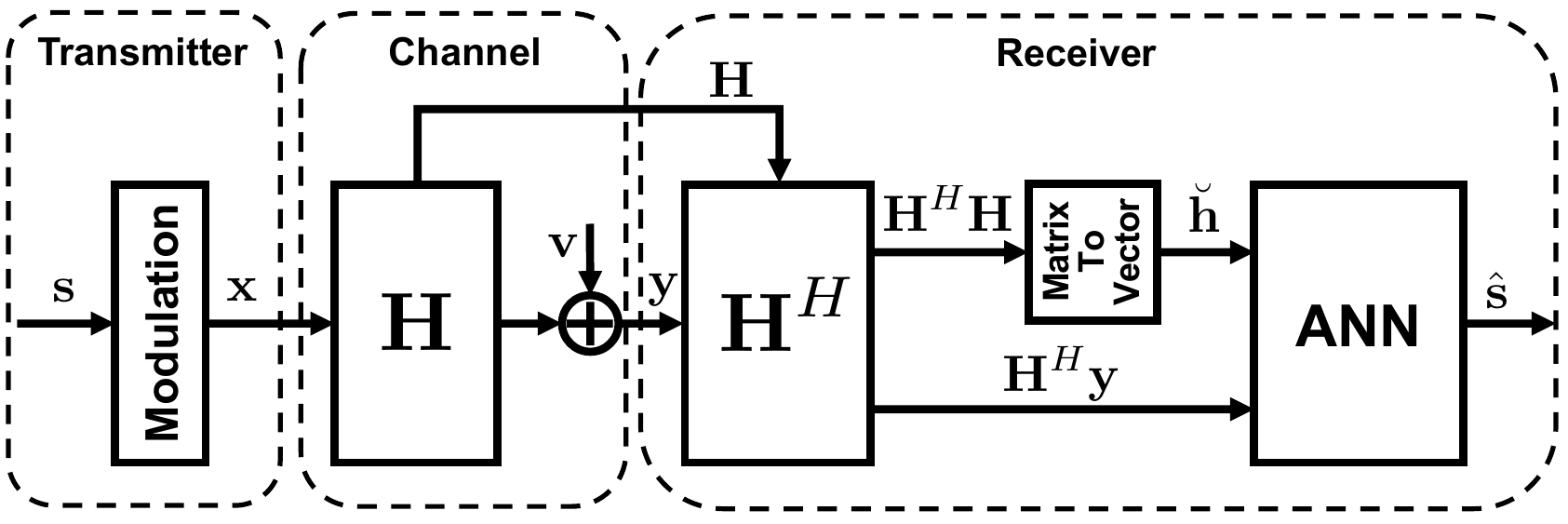}
\caption{Block diagram of the \gls{ann}-assisted \gls{mimo} detection.}\label{fig1}
\end{center}
\end{figure}

\gls{ml} signal detection faces great challenges when the \gls{mimo} channel matrix is randomly time-varying; as in this case, the set of possibly received signals becomes infinite. More seriously, the randomness of \gls{mimo} channel will result in the channel ambiguity, i.e., the receiver's observation $\mathbf{y}$ might correspond to various combinations of the channel matrix $\mathbf{H}$ and the transmitted signal block $\mathbf{x}$ even in the noiseless case. In this case, \gls{ml} is not able to conduct signal classification since the bijection between $\mathbf{y}$ and $\mathbf{x}$ is no longer hold. Theoretically, the channel ambiguity can be resolved by feeding the machine with the full channel knowledge, i.e., the input to the \gls{ann}-assisted \gls{mimo} receiver consists of the received signal block $\mathbf{y}$ as well as the channel matrix $\mathbf{H}$ or more precisely its vector-equivalent form $\breve{\mathbf{h}}$, which is often called the data-driven approach \cite{doi:10.1089/big.2013.1508}. However, the dimension of the $\mathbf{H}$-defined training input grows much faster than the $\mathbf{y}$-defined training input, and this could result in inefficient learning at the \gls{ann} training stage \cite{TCOM}. In this regard, the model-driven approach demonstrates remarkable advantages by replacing the received signal block $\mathbf{y}$ with its \gls{mf} equalized version $\mathbf{H}^H\mathbf{y}$ and the channel matrix $\mathbf{H}$ with the corresponding version $\mathbf{H}^H\mathbf{H}$; please see the block diagram of the \gls{ann}-assisted \gls{mimo} detection in \figref{fig1}. By such means, the growth rate for both inputs is largely scaled down. 

The \gls{ann} employed here is a fully-connected feedforward neural network. The input to the \gls{ann} is $\mathbf{c}_0= [ \breve{\mathbf{h}}^T,(\mathbf{H}^H\mathbf{y})^T]^T$, where $\breve{\mathbf{h}}$ is obtained by reshaping the matrix $\mathbf{H}^H\mathbf{H}$ into a vector. Assume the entire network consists of $K$ layers, \gls{ml} algorithm conducts the output through $K$ iterative steps. Denote $\matc{\omega}_k=\left \{\mathbf{W}_k, \mathbf{b}_k\right \}$ to the learning parameters of the $k$th layer, where $\mathbf{W}_k$ and $\mathbf{b}_k$ stand for the weight and bias, respectively. The stochastic objective function for the $k$th layer can be described by
\begin{equation}\label{eq03}
\mathbf{c}_{k} = \sigma _k(\mathbf{W}_{k}\mathbf{c}_{k-1}+\mathbf{b}_k),~_{k=1,2,...K}
\end{equation}
where $\sigma _k(\cdot)$ is activation function, and $\mathbf{c}_k$ is the output of $k$th layer. The most commonly used activation function for hidden layer is called rectified linear unit (ReLU), which performs a threshold operation to each element of the input. Meanwhile, the activation function of output layer is decided by the referenced training target. In \gls{mimo} signal detection, the supervised training target is the original information bit $\mathbf{s}$, which is either 0 or 1. Therefore, standard logistic function (i.e. Sigmoid) is employed, which returns a value monotonically increasing from 0 to 1. By making hard decision on the Sigmoid output, we are able to obtain the ANN estimate $\hat{\mathbf{s}}$ of the original information bits. Prior to the use of  \gls{ann} for \gls{mimo} signal detection, the \gls{ann} is trained with the aim of minimizing the following objective function
\begin{equation}\label{eq04}
\mathbb{L}(\matc{\omega}_k)=\frac{1}{S}\sum_{i=1}^{S}\mathcal{L}(\mathbf{s}_i,\hat{\mathbf{s}}_i),~_{\forall{k}}
\end{equation}
by adjusting learning parameters $\matc{\omega}_k$, where $\mathcal{L}(\cdot)$ stands for loss function, and $S$ for the size of the set which contains all input-output training pairs. Moreover, the most popular algorithm to find good set of $\matc{\omega}_k,_{\forall{k}}$ is called \gls{sgd}, which can be mathematically described by the following function
\begin{equation}\label{eq05}
\matc{\omega}_k = \matc{\omega}_k - \eta \nabla f(\matc{\omega}_k),~_{\forall{k}}
\end{equation}
where $\eta > 0$ stands for learning rate, and $\nabla f(\matc{\omega})$ for the stochastic gradient.

\subsection{The Channel Over-Training Problem}\label{sec2c}
\begin{defn}[channel over-training]\label{defn1}
It is called channel over-training problem in \gls{ann}-assisted \gls{mimo} signal detection, when an \gls{ann} optimized for a specific channel model (e.g. Rayleigh fading) is too sub-optimum or even unsuitable for other channel models.
\end{defn}

The channel over-training problem significantly limited the application of \gls{ann}-assisted \gls{mimo} detection in real practice. Potential solution towards this problem is to contentiously training the new channel model on the previous trained ANN. However, it might cause catastrophic forgetting problem, i.e., \gls{ann} has a tendency to forget previously learned knowledge when they get trained on new tasks \cite{MCCLOSKEY1989109,Smith:2017:FML:3294996.3295196}. This is reasonable since there is a shifting of input distribution across different tasks. In other words, the conventional optimization techniques converges to radically different solutions when different tasks with less common factors or structures in the input are presented to the \gls{ann}. To tackle this issue, continual learning has been proposed, which aims to inject new knowledge without forgetting previously learned knowledge. The learning process of different tasks are operated sequentially in time series. Notable continual learning approaches include:
\subsubsection{Fine-tuning \cite{DBLP:journals/corr/GirshickDDM13}} modifies the parameters of an existing \gls{ann} for the new tasks by extending the output layer with randomly initialized neurons. For the new tasks, learning rate is set to a relatively smaller value in order to mitigate its impact on the previously learned task.
\subsubsection{Network extending \cite{2019arXiv190808017T}} adds extra neurons on each layer to prevent the loss of previously learned knowledge while new discriminative features come in. The disadvantage of this method lies in the substantially increasing number of parameters in the \gls{ann}, thereby results in a performance decrease compares with the fine-tuning algorithm.
\subsubsection{Super neural network \cite{DBLP:journals/corr/FernandoBBZHRPW17}}  partitions a giant network into a number of sub-networks, with each optimized for a specific training set/task.

In addition to continual learning, a \gls{ml} paradigm called multi-task learning \cite{DBLP:journals/corr/ZhangY17aa} is able to accomplish largely the same thing. The idea lies in the use of common features contained in multiple related tasks to help improve the generalization performance of all tasks. Therefore, the learning process of different tasks are operated simultaneously in time domain.

Despite already remarkable achievements, the current state-of-the-art approaches are mainly designed for the conventional \gls{ml} applications, such as image processing and speech recognition. However, they are not cost-effective to handle wireless communication channels that are random, continuous and dynamic in their states. There is a need of a cost-effective solution that solve the channel over-training problem without complicating the \gls{ann} architecture. This motivates the development of the \gls{osgd} algorithm in this paper.

\section{The Orthogonal-SGD algorithm}\label{sec3}
The basic idea of \gls{osgd} lies in the discovery and exploitation of the orthogonality of the training samples between the current training epoch and previous training epochs. Specifically, the \gls{osgd} algorithm does not update the neural network simply based upon the current training input. Instead, it discovers the correlation between the current training samples and the historical training data. By such means, the \gls{ann} can understand and memorize uncorrelated components between different training data set. Please see {\bf Algorithm 1} for the pseudo-code of the proposed \gls{osgd} algorithm. 

\begin{algorithm}[!t]
\setstretch{1.12}
\caption{The proposed \gls{osgd} algorithm together with clipping-rate learning algorithm for stochastic optimization in \gls{ann}-assisted \gls{mimo} signal detection. Recommended parameters are $\eta_i=0.001$, $\eta_l=10^{-5}$, $\lambda = 1$, $\beta=100$, $\beta_1=0.9$, $\beta_2 = 0.999$, $\epsilon = 10^{-8}$.}%
\SetAlgoLined
	\KwIn{$\eta_i$: initial learning rate}
	\KwIn{$\eta_l$: lower bound learning rate}
	\KwIn{$\lambda$: forgetting factor}
	\KwIn{$\epsilon$: compensation factor}
	\KwIn{$\mathbf{c}_{k-1}^t$: input to the $k$th layer at time slot $t$}	
    \KwIn{$f(\matc{\omega}_{k})$: stochastic objective function}
    \KwIn{$\beta_1, \beta_2 \in[0,1)$:  exponential decay rates for the moment estimates}
         {\bf Initialization:} $\matc{\omega}_{k}$, $\mathbf{\Psi} _{k}=\mathbf{I}_k/\beta$, $\mathbf{m}_{k} = \mathbf{0}$, $\mathbf{v}_{k} = \mathbf{0}$,\\ 		           \qquad \qquad  \qquad $ $ $t=0$\\
   		 \While{$\matc{\omega}_{k}$ not converged}{
   		 	$t = t + 1$ \\
        		\For{$k=1$ {\bf to} $K$}{
        		$\mathbf{p}_{k} = \mathbf{\Psi}_{k}\mathbf{c}_{k-1}^t/(\lambda+\mathbf{c}_{k-1}^{t\quad T}\mathbf{\Psi} _{k}\mathbf{c}_{k-1}^t)$ \\
        		$\mathbf{\Psi} _{k}  = \lambda^{-1}\mathbf{\Psi} _{k} - \lambda^{-1} \mathbf{p}_{k} \mathbf{c}_{k-1}^{t \quad T} \mathbf{\Psi} _{k}$ \\
        		$\mathbf{g}_{k} = \nabla f(\matc{\omega}_{k}) \cdot \mathbf{\Psi}_k^T$ \\
        		$\mathbf{m}_{k} = \beta_1\mathbf{m}_{k} + (1-\beta_1) \cdot \mathbf{g}_{k}$ \\
        		$\mathbf{v}_{k}  = \beta_2\mathbf{v}_{k} + (1-\beta_2) \cdot \mathbf{g}_{k}^2$ \\
        		$\hat{\mathbf{m}}_{k}  = \mathbf{m}_{k} /(1-\beta_1^t)$\\
        		$\hat{\mathbf{v}}_{k}  = \mathbf{v}_{k} /(1-\beta_2^t)$\\
        		$\eta_t = \max(\eta_i/\sqrt[4]{t},\eta_l)$ \\
        		$\matc{\omega}_{k} = \matc{\omega}_{k} - \eta_t \cdot \hat{\mathbf{m}}_{k} /(\epsilon+\sqrt{\hat{\mathbf{v}}_{k} })$\\      		
        		}
        }
     \KwOut{$\matc{\omega}_{k},~_{\forall k}$}
\label{algo:solution}
\end{algorithm}

Denote $f(\matc{\omega}_{k})$ to a noisy stochastic objective function with respect to the parameter $\matc{\omega}_{k}$, as we have introduced in \eqref{eq04}, where the subscript
$[\cdot]_k$ stands for the layer number. The aim of \gls{ann} training process is to minimize the expected value of the objective function $\mathbb{E}[f(\matc{\omega}_{k})]$ by adjusting the network parameters $\matc{\omega}_{k}$. 

In \gls{osgd}, the first step is to calculate the correlation between the current input $\mathbf{c}_{k-1}^t$ and all previous training data. Mathematically, it can be done by forming all previous input training data as a matrix $\mathbf{A}$ and calculating the direction orthogonal to the space of matrix $\mathbf{A}$ by
\begin{equation}\label{eq06}
\mathbf{\Psi}_k = \mathbf{I} - \mathbf{A}(\mathbf{A}^T\mathbf{A} + \alpha \mathbf{I})^{-1}\mathbf{A}^T
\end{equation}
where $[\cdot]^{-1}$ stands for the vector/matrix inversion, and $\alpha$ is a relatively small constant. The direction-modified gradient is then determined by 
\begin{equation}\label{eq07}
\nabla f(\matc{\omega}_{k})' = \nabla f(\matc{\omega}_{k}) \cdot {\mathbf{\Psi} _{k}}
\end{equation}
where $\nabla f(\matc{\omega}_{k})$ is the gradient obtained by the conventional \gls{sgd} algorithm. However, the computational complexity and memory requirement for such an approach continue to increase over time, which makes it unsuitable for real practise. Inspired by the \gls{rls} algorithm in adaptive filter theory \cite{Haykin:1996:AFT:230061}, we improve the \gls{osgd} algorithm by considering each time step as an independent task and updating $\mathbf{\Psi} _{k}$ through an iterative manner
\begin{align}\label{eq08}
\mathbf{p}_{k} &= \mathbf{\Psi} _{k}\mathbf{c}_{k-1}^t/(\lambda+\mathbf{c}_{k-1}^{t\quad T}\mathbf{\Psi} _{k}\mathbf{c}_{k-1}^t) \nonumber \\
\mathbf{\Psi} _{k}  &= \lambda^{-1}\mathbf{\Psi} _{k} - \lambda^{-1} \mathbf{p}_{k} \mathbf{c}_{k-1}^{t \quad T} \mathbf{\Psi} _{k}
\end{align}
where $\mathbf{c}_{k-1}^t$ is the input to the $k$th layer at time step $t$, $\lambda$ is the forgetting factor, and $\mathbf{\Psi}_{k}$ is initialized as $\mathbf{\Psi} _{k}=\mathbf{I}_k/\beta$, where $\beta$ is a constant number. It is perhaps worth noting that $\mathbf{\Psi} _{k}$ is layer-specific, each layer has to compute it independently based on the input $\mathbf{c}_{k-1}^t$. By such means, neural network does not need to store all previous training data, instead only the previous gradient projection matrix $\mathbf{\Psi}_{k}$ is needed.

The second step is to update the network parameters by employing the first-order gradient-based optimization approach. The most commonly used algorithm is Adam \cite{Kingma2014AdamAM}. However, researchers have recently found that Adam can fail to converge to an optimal solution even in simple one-dimensional convex settings \cite{DBLP:journals/corr/abs-1904-09237}. Meanwhile, we also observed the convergence difficulties in the training process of \gls{ann}-assisted \gls{mimo} signal detection. To tackle this issue, we designed a low-complexity clipping-rate optimization algorithm which takes the merits of the original Adam algorithm and mitigates the non-convergence problem particularly at the end of the training stage. The learning rate clipping function can be described as
\begin{equation}\label{eq09}
\eta_t = \max(\eta_i/\sqrt[4]{t},\eta_l)
\end{equation}
where $\eta_i$ is the initialized learning rate and $\eta_l$ is the lower bound of the learning rate. In this paper, we set $\eta_i = 0.001$ and $\eta_l = 10^{-5}$, as the above configuration is found to provide the best performance in our computer simulations.

\section{Simulation results and evaluation}
This section presents the experimental results and related analysis. The training data set and experimental settings are firstly introduced, followed by three experiments which aim to demonstrate our hypotheses in previous sections.

\subsection{Data Sets and Experimental Setting}
In conventional \gls{ml} applications, such as image processing and speech recognition, different algorithms are evaluated under common benchmarks or data sets. However, wireless communication system normally deals with artificially manufactured signals which can be accurately generated. Thereby, we prefer to define the communication system model instead of providing specific training data sets. 

In order to explore the channel over-training problem inherent in \gls{ann}-assisted \gls{mimo} signal detection, the performance of different algorithms are evaluated under multiple Rician fading channels with different values of $K$, where $K$ denotes the ratio between the power in the direct path (i.e. line-of-sight) and the power in the scattered paths (i.e. non-line-of-sight). The probability density function (PDF) of Rician distribution can be described by
\begin{equation}\label{eq10}
f(x|\nu,\sigma)=\frac{x}{\sigma^2}\exp \bigg( -\frac{x^2+\nu^2}{2\sigma^2}  \bigg)I_0\Big( \frac{x\nu}{\sigma^2}  \Big)
\end{equation}
with $\nu^2 = K\Omega/(K+1)$ and $\sigma^2=\Omega/(2(K+1))$, where $I_0(\cdot)$ is the modified Bessel function of the first kind with order zero, and $\Omega$ is the total power from both paths which is normalized to one. Moreover, we assume that the value of $K$ randomly varies in the range of [0, 5]. When $K$ equals to 0, the channel model becomes Rayleigh fading which has been widely used in state-of-the-art. 

Recall the \gls{ann} architecture in \figref{fig1}, the transmitted signal block $\mathbf{x}$ is the modulated information bits $\mathbf{s}$. Channel coding is not considered in our simulations, but it can be straightforwardly implemented on the proposed architecture with no performance penalty. At the receiver side, the signal block is firstly proposed by the conventional \gls{mf} equalizer; and then, together with the vector-form channel knowledge $\breve{\mathbf{h}}$, serves as the input to the \gls{ann}. It is perhaps worth noting that the received signal as well as the channel knowledge are complex-valued, however most of the \gls{ml} algorithms are designed by using real-valued numbers. To facilitate the operation of \gls{ann}, the complex-valued signal block is represented by the real-valued block with double the size (i.e. real and imaginary parts are concatenated). As far as the supervised learning is concerned, each training input should be paired with a supervisory output. In this paper, the referenced training target is set to be the original information block $\mathbf{s}$ as we have introduced in Section II-B. The detailed layout of \gls{ann} for \gls{mimo} signal detection can be found in Table I.

All the experiments are run on a Dell PowerEdge R730 2x 8-Core E5-2667v4 Server, and implemented in MATLAB.

\begin{table}[h]
\footnotesize
\centering
\caption{Layout of the \gls{ann} used in all experiments}\label{T1}
\renewcommand{\arraystretch}{1.2}
\begin{tabular}{p{4cm}|p{3cm}}
\hline
\textbf{Layer} &\textbf{Output dimension} \\ \hline
Input  & $2M (M+1)$ \\ 
Dense + ReLU & $512$   \\ 
Dense + ReLU & $256$   \\
Dense + ReLU & $128$   \\ 
Dense + Sigmoid & $M\log L$   \\ \hline
\end{tabular}
\end{table}

\begin{figure}[t!]
\centering
\includegraphics[width=3in]{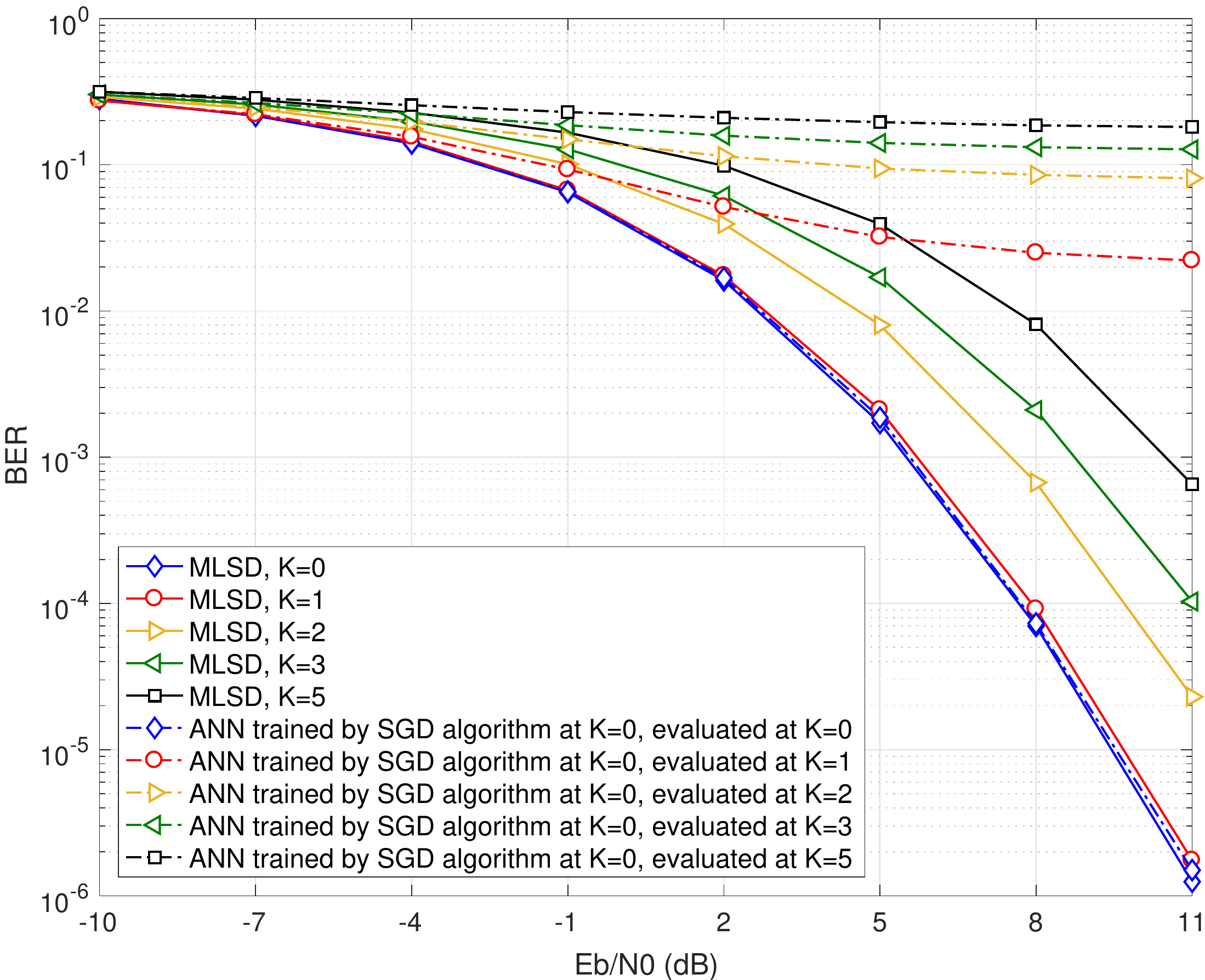}
\caption{\gls{ber} as a function of $E_b/N_0$ for \gls{ann}-assisted \gls{mimo} signal detection. The \gls{ann} is trained for Rayleigh fading channel (i.e. $K=0$), and evaluated under various channel models.}
\label{fig2}
\end{figure}

\subsection{Simulation and Performance Evaluation}
Our computer simulation are structured into three experiments. The first experiment aims to demonstrate our hypothesis on the existence of the channel over-training problem in \gls{ann}-assisted \gls{mimo} signal detection. Experiment 2 and 3 evaluate the performance of the conventional \gls{sgd} algorithm and the proposed \gls{osgd} algorithm by training multiple tasks sequentially and simultaneously in time domain, respectively. The size of the \gls{mimo} system is 4-by-8, and QPSK modulation is considered at the transmitter side. The key metric utilized for performance comparison is the average \gls{ber} over sufficient Monte-Carlo trails of multiple block fading channels. Moreover, the \gls{snr} is defined as the average received information bit-energy to noise ratio per receive antenna (i.e. $E_b/N_0$).

\subsubsection*{Experiment 1}
In this experiment, an \gls{ann}-assisted \gls{mimo} receiver optimized for Rayleigh fading channel (i.e. $K=0$) is evaluated under multiple other channel models. The aim of this experiment is to demonstrate the existence of channel over-training problem in \gls{ann}-assisted \gls{mimo} signal detection. Moreover, the training is operated at $E_b/N_0 = 8 $ dB with a mini-batch size of 500; as the above configurations are found to provide the best performance.

\figref{fig2} shows the average \gls{ber} performance of the \gls{ann}-assisted \gls{mimo} receiver trained by \gls{sgd} algorithm. The baseline for performance comparison is the optimum \gls{mlsd}. It is shown that the \gls{ann}-assisted \gls{mimo} receiver achieves near-optimum performance under trained channel model (i.e. $K=0$); the performance gap to the \gls{mlsd} is almost negligible. However, the detection performance significantly decreased when other channel models are considered (i.e. $K=1,2,3,5$). The performance gap between the \gls{ann}-assisted \gls{mimo} receiver and \gls{mlsd} is more than 10 dB at high \gls{snr} regime. The above phenomena coincides with our hypothesis that the channel over-training problem exists in the \gls{ann}-assisted \gls{mimo} signal detection.

\subsubsection*{Experiment 2}
\begin{figure}[t!]
\centering
\includegraphics[width=3in]{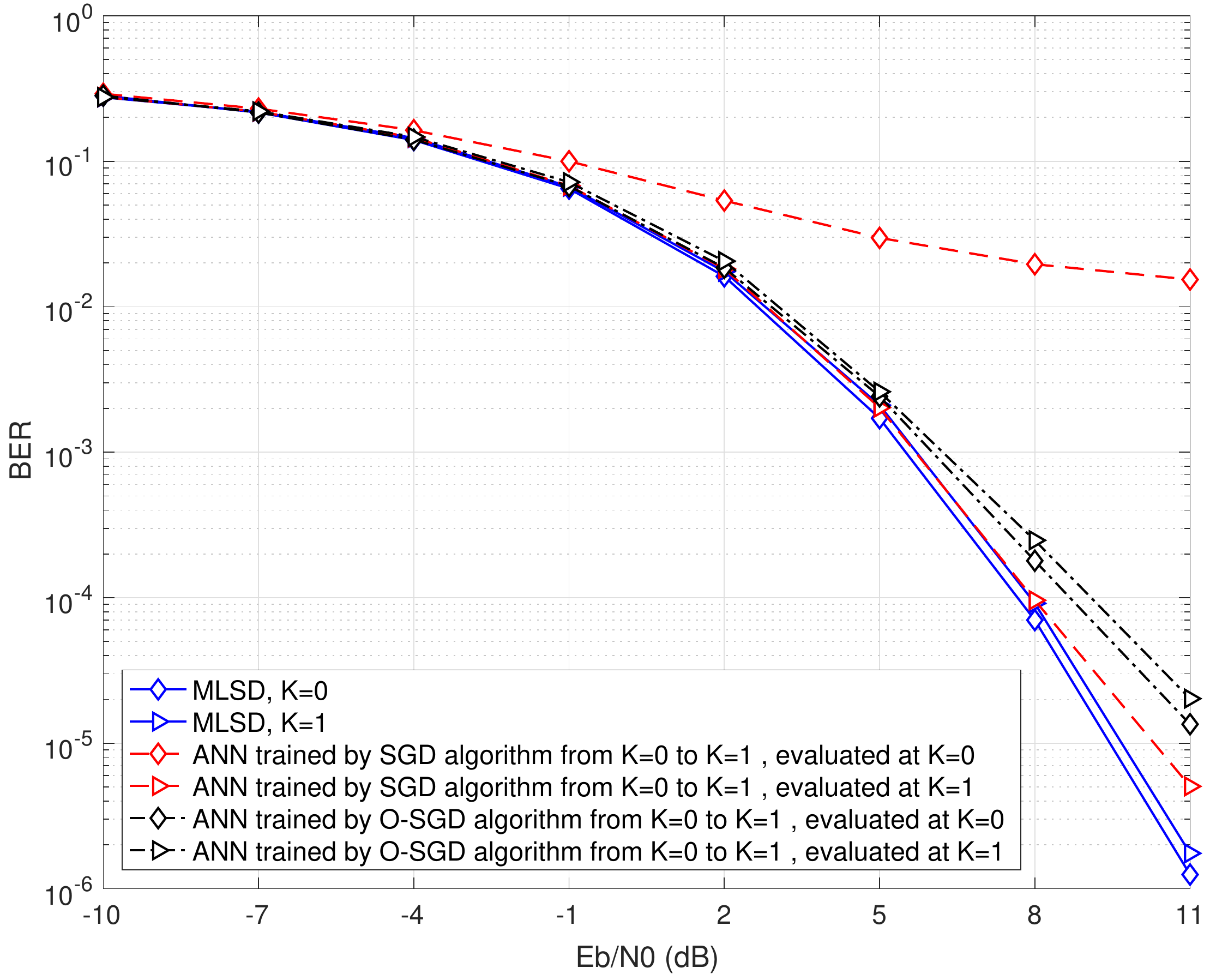}
\caption{\gls{ber} as a function of $E_b/N_0$ for \gls{ann}-assisted \gls{mimo} signal detection. The \gls{ann} is firstly trained under Rayleigh fading channel (i.e. $K=0$), and then trained for Rician fading channel (i.e. $K=1$), and evaluated under both channel models.}
\label{fig3}
\end{figure}

In this experiment, \gls{ann}-assisted \gls{mimo} receiver is firstly trained under Rayleigh fading channel (i.e. $K=0$) by using either \gls{sgd} or \gls{osgd} algorithm. After training converge, a new training task (i.e Rician fading channel with $k=1$) is operated on the previously trained \gls{ann}. The detection performance is evaluated under both channel models. The training $E_b/N_0$ is set at 8 dB and the size of the mini-batch is 500.

\figref{fig3} shows the average \gls{ber} performance of the \gls{ann}-assisted \gls{mimo} receiver trained by either \gls{sgd} or \gls{osgd} algorithm. The baseline for performance comparison is the optimum \gls{mlsd}. It is shown that \gls{sgd} algorithm fails to remember the previously learned knowledge, as the BER performance for the first task (i.e. $K=0$) is around 9 dB away from \gls{mlsd} at \gls{ber} of $10^{-2}$. Conversely, it achieves a near-optimum performance for the second channel model (i.e. $K=1$). The gap between \gls{mlsd} and \gls{sgd} is less than 1 dB at \gls{ber} of $10^{-5}$. On the other hand, the proposed \gls{osgd} algorithm shows promising learning capabilities for both tasks. The performance gaps to the optimum \gls{mlsd} are 1.2 dB and 1.4 dB at \gls{ber} of $10^{-4}$ for $K=0$ and $K=1$, respectively. It is also observed that the first task slightly outperforms the second task by around 0.5 dB at \gls{ber} of $10^{-4}$. The above phenomena demonstrates our hypothesis in Section II-C that the proposed \gls{osgd} algorithm is able to mitigate the channel over-training problem, but \gls{sgd} can not.

\subsubsection*{Experiment 3}

\begin{figure}[t!]
\centering
\includegraphics[width=3in]{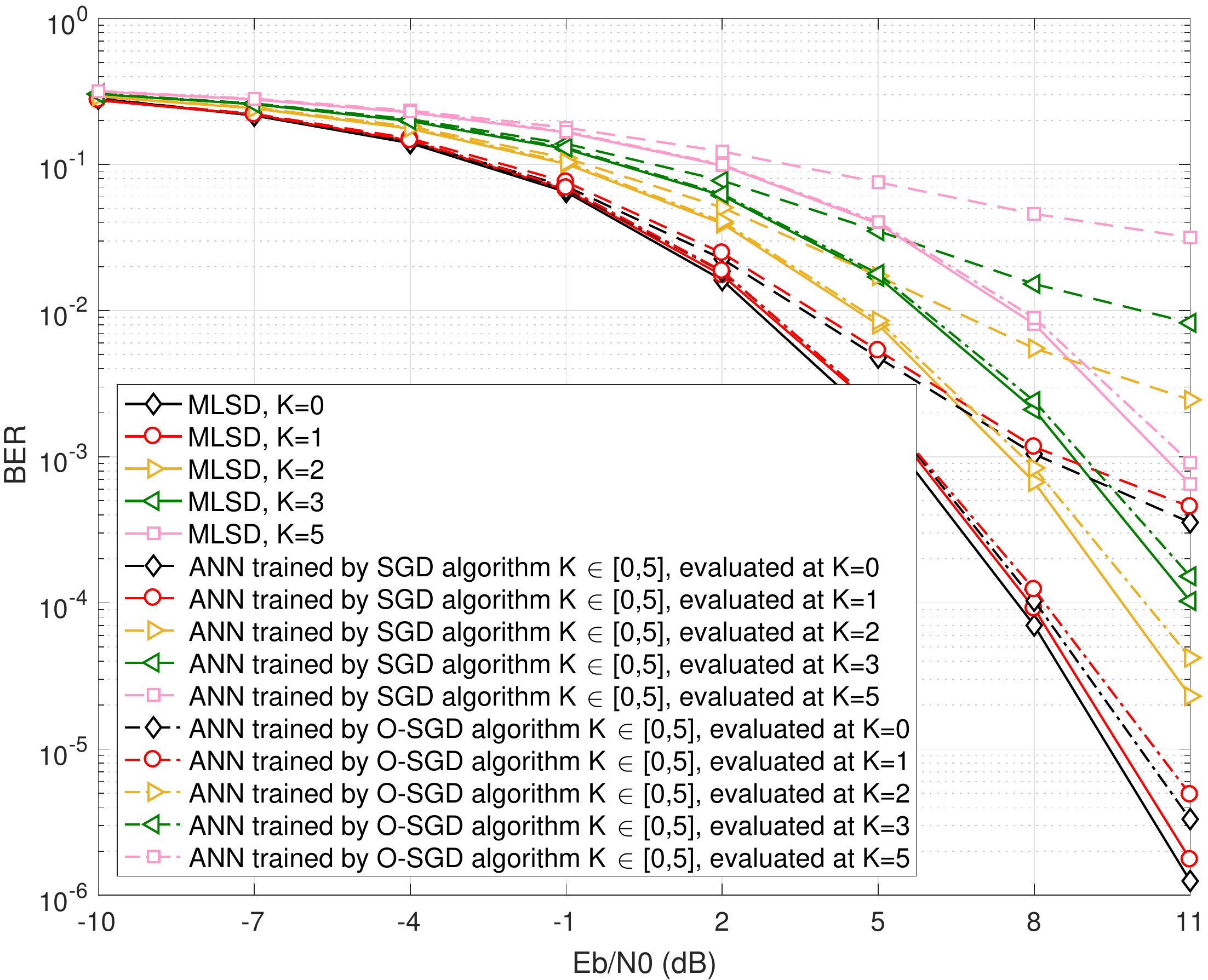}
\caption{\gls{ber} as a function of $E_b/N_0$ for \gls{ann}-assisted \gls{mimo} signal detection. The \gls{ann} is trained under a mix of multiple RIcian fading channel models in the range of $K \in [0,5]$, and evaluated under a number of selected channel models.}
\label{fig4}
\end{figure}

In this experiment, \gls{ann}-assisted \gls{mimo} receiver is trained under a mix of multiple Rician fading channel models with $K$ randomly varies in the range of [0, 5]  by using either \gls{sgd} or \gls{osgd} algorithm. Besides, the training settings remain unchanged as we introduced in the previous experiments.

\figref{fig4} shows the average \gls{ber} performance of the \gls{ann}-assisted \gls{mimo} receiver trained by either \gls{sgd} or \gls{osgd} algorithm. The baseline for performance comparison is the optimum \gls{mlsd}. It is shown that the proposed \gls{osgd} algorithm achieves promising detection performance under all the selected channel models (i.e. $K=0,1,2,3,5$). The gap between \gls{mlsd} and \gls{osgd} is less than 1 dB. By contraries, \gls{sgd} fails to conduct a good signal detection specifically at high \gls{snr} regime. The gap to the optimum \gls{mlsd} is around 2.5 dB at \gls{ber} of $10^{-3}$ for $K=0$ and $K=1$, and more than 5 dB for the other three channel models. This phenomenon indicates that the proposed \gls{osgd} algorithm is able to achieve promising performance when multiple tasks are learned simultaneously.

\section{Conclusion}
In this paper, a novel \gls{osgd} algorithm has been introduced to handle the channel over-training problem inherent in the \gls{ann}-assisted \gls{mimo} signal detection. It has been shown that \gls{osgd} can discover the orthogonality between the current training epoch and previous training epochs, and update the neural network by exploring the uncorrelated components among different training tasks. Simulation results have shown that the proposed \gls{osgd} algorithm significantly outperforms the conventional \gls{sgd} algorithm under multiple channel models. 

{
\section*{Acknowledgement}
The work was supported in part by European Commission under the framework of the Horizon2020 5G-Drive project, and in part by 5G Innovation Centre (5GIC) HEFEC grant.
}

\ifCLASSOPTIONcaptionsoff
  \newpage
\fi

\balance
\bibliographystyle{IEEEtran}
\bibliography{ref.bib}
\end{document}